\documentclass[manuscript]{aastex}
\usepackage{graphicx}

\shorttitle{A Turbulent Interstellar Medium Origin of the Binary Period Distribution}
\shortauthors{Fisher, R. T.}

\begin{document}


\title{A Turbulent Interstellar Medium Origin of the Binary Period Distribution}  

\author{Robert T. Fisher}
\email{bobf@astron.berkeley.edu}
\affil {Lawrence Livermore National Laboratory, Mail Code L-023, 7000 East Avenue, Livermore Ca. 94550}


\begin{abstract}

In this paper, we present a semi-empirical model of isolated binary star formation. This model includes the effects of turbulence in the initial state of the gas, and has binary orbital parameters consistent with observation. Our fundamental assumption is that the angular momenta of binary star systems
 is directly related to the net angular momentum induced by turbulence in parent
 molecular cloud cores.      
The primary results of this model are as follows. (i) A quantitative prediction of the initial width of the binary period distribution ($\sigma_{\log {P}_d} = 1.6 - 2.1$ for a star formation efficiency in the range $\epsilon_* = 0.1 - 0.9$). (ii)  A robust, negative anticorrelation of binary period and mass ratio. (iii) A robust, positive correlation of binary period and eccentricity. (iv) A robust prediction that the binary separation of low-mass systems should be more closely separated than those of solar-mass or larger. These predictions are in good agreement with observations of PMS binary systems with periods $P > 10^3$ d, which account for the majority of all binaries.

We conclude with a brief discussion of the implications of our results for observational and theoretical studies of multiple star formation.
 
\end{abstract}



\section {Introduction}

To fully elucidate the mechanisms underlying the origin of binary systems, it is crucial to compare theory with observations of the field and of nearby star-forming regions. However, a myriad of complex physical effects have been advanced by theorists to account for the formation and evolution of binary and multiple star systems, including fragmentation \citep {hoyle53, inutsukamiyama92}, turbulence \citep {kleinetal01, bateetal02}, ideal magnetohydrodynamics (MHD) \citep {gallietal01}, non-ideal MHD, including ambipolar diffusion \citep {mouschovias77, boss00}, radiative transfer \citep {bossetal00}, dust physics \citep{whitworthetal98}, tidal torquing \citep {larson02},  capture \citep {clarkepringle91a},  competitive accretion \citep {bonnelletal97}, and accretion disk processes \citep {laughlinandbodenheimer94}. However, to date, no simulation has included all of these varied physical effects, and few have been run to a substantial evolutionary state. Moreover, as other authors have pointed out \citep {larson01}, once one includes the effect of turbulence, star formation is inherently a stochastic process, implying that many such simulations need to be performed before their results can be meaningfully compared to observation.

In this paper, we present a simplified, semi-empirical model of isolated binary star formation. This model includes the effects of turbulence in the initial state of the gas, and has binary orbital parameters consistent with observation. Our fundamental assumption is that {\it the angular momenta of binary star systems is directly related to the net angular momentum induced by turbulence in parent molecular cloud cores.} We do not require that the binary angular momentum be equal to that of its parent core; indeed, we shall find that angular momentum loss from the initial core is crucial in determining the median binary period, as previous authors suggested \citep {simon92, bodenheimer95}.  While we do not in this work determine the mechanism underlying the loss of angular momentum, we find that we can describe a number of observed properties simply by specifying that the efficiency of transfer of mass and angular momentum from the initial gaseous core to the final binary are both constant. This model therefore provides an excellent framework with which to examine the importance of turbulence in the context of isolated star formation, and to explore the role which the stochastic nature of the initial turbulent state has on the statistical properties of binary systems. While it is certainly true that a number of other physical effects other than turbulence and angular momentum dissipation may play a role in the star formation process, our model can elucidate their relative importance by serving as a strawman against which other results can be compared. 

The paper begins with an overview of observational evidence (\S \ref {observationalevidence}), including both pre-stellar molecular cloud core (\S \ref {prestellarcore}) and binary (\S \ref {binaryprop}) properties. Next, we describe our pre-stellar core model in \S \ref {model}. In \S \ref {methodology}, we present the methodology which connects the pre-stellar core model with the binary properties. The results of this model are detailed in \S \ref {results}. We conclude the main body of the paper with a discussion of the implications of this semi-empirical model for observational and theoretical studies of star formation (\S \ref {discussion}). Lastly, appendix A includes an elementary derivation of characteristic binary scales, in terms of their orbital parameters.  

\section {Observational Evidence}
\label {observationalevidence}

\subsection {Pre-Stellar Core Properties}
\label {prestellarcore}

\subsubsection {Density Structure}
\label {densitystructure}

\begin{sloppypar}
The observations of \citet {motteandre01} consisted of  a complete 1.3 mm survey of both pre-stellar cores and protostellar envelopes using the IRAM 30m telescope and the MPIfR bolometer array (MAMBO). Their observations resolved structures from 1,500 AU to 15,000 AU. \cite {motteandre01} concluded that their observations of pre-stellar core backgrounds were flatter than predicted by a singular isothermal sphere model \citep {shu77}, but were consistent with Bonnor-Ebert spheres \citep {bonnor57, ebert55}. 
\end{sloppypar}

\subsubsection {Velocity Structure on the Molecular Cloud Core Scale}
\label {velocitystructure}

In 1981, in an extension of an earlier paper \citep {larson79}  focusing on the large-scale structure of the interstellar medium,  \citet {larson81}  published a seminal paper on the internal velocity dispersion in molecular clouds. Using data already published in the literature in a wide variety of studies, he accumulated about 50 data points for a number of different star forming regions, on scales ranging from $10^{-1}$ to $10^2$ pc. He established scaling laws for both the mean density and the total internal velocity dispersion over projected lengthscales on the sky. Plotting the total internal velocity dispersion (sum of thermal and non-thermal linewidths) versus the maximum projected linear size of the region, he found that the internal velocity dispersion obeyed a power law of the form
\begin{equation}
\sigma_{1D} (\rm {km\ s^{-1} }) = 1.10\ L (\rm {pc})^{0.38}
\label{larsondisp}
\end{equation}
Here we have designated the one-dimensional linewidth dispersion, as inferred from velocities along the line of sight by $\sigma_{1D}$, to properly distinguish it from the fully 3D velocity dispersion. 

Significantly, the power law nature of the turbulent linewidth-size scaling laws supports the premise, which Larson originally suggested, that over a wide range of scales, the observed interstellar turbulence is part of a scale-free hierarchy of turbulent eddies. This scale-free cascade should extend all the way down to about the characteristic scale where such eddies are damped -- either the minimum Jeans mass in the cloud \citep {larson95} or the ion-neutral damping length \citep {myerslazarian98}, depending on whether thermal or magnetic damping effects dominate, which in turn depends on the relative strength of the magnetic field and the ionization state of the gas.

Subsequent authors investigated the linewidth-size scaling relationship in more detail for molecular cloud cores using more precise techniques. In particular, Larson's original study contained only 4 cores observed with rather poor angular resolution, and in the absence of actual data, made blanket assumptions regarding the kinetic temperature of the gas used in computing the local soundspeed.  \citet {leungetal82} and \citet {myers83} re-investigated Larson's linewidth- size relationship and found the exponent closer to $1/2$, rather than $1/3$. In the most extensive study of molecular cloud cores to date,  \citet {jijinamyersadams99} studied 264 cores mapped in NH$_3$ in a wide variety of regions and environmental conditions. They conclude that the linewidth-size relationship for all cores has an exponent of $.63 \pm .10$, though a slight difference of low statistical significance existed between subsamples with embedded IRAS sources (exponent $.49 \pm .12$) and without (exponent $.83 \pm .18$).  Moreover, \citet {jijinamyersadams99} found a significant variation in the median value of the non-thermal linewidth $\Delta v_{NT}$ across regions : from a value of 0.22 km/s in Taurus (thermal linewidth median $\Delta v_T = .44$ km/s), to a non-thermal linewidth $\Delta v_{NT}$  median value of .86 km/s in Orion A (thermal linewidth median $\Delta v_T$ = .58 km/s). Over all regions, cores without IRAS embedded sources or clusters had  $\Delta v_{NT} / \Delta v_T = .65^ {+ .35}_{- .25}$ (within one quartile of the median). Cores without clusters, but with IRAS embedded sources had $\Delta v_{NT} / \Delta v_T =  1.0^ {+ .40}_{- .30}$.  Cores with clusters had higher non-thermal linewidths still.

We note that the most complete observations to date suggest that even in low-mass cores, the level of turbulent support is comparable to that of thermal support, although there are large variations both within and across individual star-forming regions.\footnote{Although these cores are often termed ``quiescent", such a designation is somewhat misleading.} Moreover, there is little evidence that the non-thermal linewidths diminish during the process of star formation; if anything, the non-thermal linewidths in cores with IRAS embedded sources is {\it higher} than in their non-IRAS counterparts. However, we note that the non-thermal linewidths measured in cores with embedded objects will in general also include infall motion towards one or more protostars in a complex flow field.  Such infall motions are extremely difficult, if not impossible, to separate from turbulent motions on the scale of the core without a full spatial mapping of the velocity field, and so it is remains unclear (from an observational standpoint) what the timescale for the dissipation of turbulence is, or indeed, even whether turbulence decays or increases during star formation.

\subsection {Binary Star Properties}
\label {binaryprop}

\subsubsection {Binarity Fraction}
Prompted by inconsistencies in multiplicity fractions in earlier surveys \citep {kuiper35, kuiper42,
 jaschekjaschek57, petrie60, jaschekgomez70}, \citet {abtandlevy76} determined the multiplicity fraction of a sample  of 135 bright field stars of types F3-G2, and found a binarity fraction (the fraction of all primaries having a detectable secondary companion) of 57\%.  Later, \citet {dm91} (DM91 hereafter) using a complete sample of 164 primary G dwarf stars with types F7-G9, found a remarkably similar binarity fraction of 58\% (although their conclusions regarding the period distribution differed substantially; see below).  

\subsubsection {Period Distribution}

DM91 demonstrated that the binary periods in a complete sample of field G dwarf stars fits a broad log-normal distribution remarkably well, with a median period of $\log \bar{P}_d = 4.8$  and a standard deviation of $\log \sigma_{P_d} = 2.3$, where $P_d$ is the period measured in days. DM91's result is substantially different than the median period $\log \bar{P}_d = 3.7$  found previously by \citet {abtandlevy76}, whose survey was biased by their use of a magnitude-limited sample.  Similar period distributions were found for field K dwarfs \citep {mayoretal92} and field M dwarfs \citep {fischerandmarcy92}, though the K dwarf sample was found to have a somewhat narrower distribution ($\log \sigma_{P_d} = 1.9$).

\citet {mathieu94} compiled data of pre-main sequence (PMS) binaries extant at that time. While the statistics of PMS binaries are not as complete as those in the field, the compiled PMS binary period distribution appears to fit a log-normal distribution similar to that of \citet {dm91}. More recently, other observers have begun to produce tentative evidence in favor of the hypothesis that the period distribution itself varies from region to region. For instance, in observations of Upper Scorpius A and B, \citet {brandnerkoehler98} argue that the period distribution median varies significantly between the two regions, with wider binaries being preferred in the low-mass star forming region. Similarly, \citet {scallyetal99} argue that a deficiency of wide ($a > 1000$ AU) binaries exists in Orion. However, such studies of individual regions and distribution tails reduce the counting statistics by necessity (some 20 systems in the case of Brandner and Koehler's study; some 3 systems in the case of Scally et al's study), and so we must consider them  with some caution.

To date, the period distribution has not been quantitatively explained by theory. \citet {mouschovias77} envisioned dense cores as non-turbulent, highly subcritical, slowly contracting regions initially endowed with angular velocity induced by galactic shear. In his picture, magnetic braking becomes inefficient at some critical density at which the core becomes supercritical, and thereafter the angular momentum of the gas is conserved during the collapse.  He invoked variations in this critical density parameter across all star-forming regions to explain the broad variation in the binary period distribution. However, it remains unclear whether this picture is quantitatively consistent with the observed period distribution.

More recently, \citet {larson02} offered the possibility that tidal interactions among binaries in a stellar cluster would tend to transfer angular momentum among the binaries, and thereby broaden the period distribution. However, \citet {kroupaburkert01} demonstrated that even under extreme conditions (stellar densities $\sim 10^6$ pc$^{-3}$), gravitational encounters were not able to explain the observed breadth of the binary period distribution. Indeed, even when a very small amount of broadening of the period distribution was obtained in their densest configurations, disruption dominated over broadening, so that the binarity fraction dropped from an initial value of unity to $\sim .2$ in about $3 \cdot 10^4$ yr, in sharp conflict with observation. Indeed, \citet {kroupaburkert01} concluded that {\it the binary period distribution was a remnant of the initial conditions of star formation, and was not due to subsequent evolution}. However,  \citet {bateetal02} found that close systems could form in a gas-rich environment through dynamical friction. A corollary of Bate's findings is that in a turbulent environment, the angular momentum dissipation rate should be sensitive to initial conditions -- as had been suggested earlier by \citet {larson01}. We will return to this point later (\S \ref {discussion}).

\subsubsection {Eccentricity Distribution}
\label {eccentricitydistr}

Another important diagnostic of binary stars is the eccentricity distribution. DM91 divided their
binary catalog into short ($P < 10^3$ d) and long ($P > 10^3$ d) samples. They found that while
the short-period binaries had a distribution which peaked at $e = .3$, long-period
binaries followed a distribution which was roughly linear ($f (e) = 2 e$), up to eccentricities 
of about $.8$, at which point the eccentricity distribution tapers off significantly.  

\citet {mathieu94}  reviewed the data extant at that time for MS and PMS binaries, and compared plots of the eccentricity $e$ versus period $P$ for each population of  binaries separately. One of his principal findings was that the eccentricities of longer period  systems (both MS and PMS) were positively correlated with the period of the system, to periods of at least 10$^3$ days. Moreover, the MS eccentricity distribution appears to be roughly consistent with the PMS eccentricity distribution. This suggests that the eccentricities of MS binaries are established by age of the PMS stage (about $10^6$ years) or earlier.

\subsection {Mass Ratio}

An accurate observational determination of the the mass ratio distribution requires a number of corrections due to observational biases. DM91 found a mass ratio distribution which was consistent with drawing both individual members of the binary system independently from the IMF, and could be fitted to a Gaussian. In particular, their mass ratio distribution exhibited no maximum towards $q = 1$. However, a later paper \citep {mazehetal92}, pointed out that once additional selection-effect corrections for spectroscopic binaries with periods less than $3 \cdot 10^3$ d are taken into account, the resultant period distribution for short period binaries is nearly flat, with a slight rise towards $q = 1$.
  
\section {Initial Molecular Cloud Core Model Formulation}
\label {model}

\subsection {Overview}

Our model for turbulent molecular cloud cores will consist of two essential elements. The first, the background density model (\S \ref {backgrounddensity}) is based upon a Bonnor-Ebert density profile of isothermal gas, with pressure support enhanced through turbulent pressure. As we will show, this density profile is consistent both with Larson's mean density relationship and with resolved observations of molecular cloud cores (\S \ref {densitystructure}).  The second element, a turbulent velocity field (\S \ref {velocityperturbations}),  is superposed upon this background density. This velocity field is consistent with Larson's linewidth-size relation (\S \ref {velocitystructure}).

In contrast to our assumptions of isothermal gasdynamics and a turbulent velocity field, other authors have have assumed that turbulence can effectively be accounted for by using a logotropic \citep {mclaughlinpudritz97} or polytropic pressure component \citep {mckeeholliman99}, and derived density profiles for their models under the assumption that gravity was balanced by the thermal and turbulent pressure of the core. Our approach differs fundamentally in that it is essentially a {\it non-equilibrium} model -- while the kinetic, thermal, and gravitational energies are in virial balance, our cores are never static. This difference is absolutely critical in the context of the current work, since we seek to tie the binary properties to the angular momentum of the turbulent model cores.

\subsection {Background Density Model}
\label {backgrounddensity}

Since our models include turbulent support, we define an effective soundspeed $c_{\rm eff}$, which includes contributions from both thermal and turbulent pressure, added in quadrature. Because pressure depends on only the 1D rms of the velocity, we include the one-dimensional turbulent support through the effective soundspeed :
\begin{equation}
c_{\rm eff}    = c_{\rm iso}  \sqrt { 1 + \left (
                       {\mathcal {M} \over \sqrt {3} }  \right)^2}
\end{equation}

It is both convenient and physically meaningful to scale the radius $r$ to a dimensionless unit $\xi = r / r_0$, where $r_0$ proportional to the Jeans length at the central density $\rho_c$ and to scale the density itself to a dimensionless  unit $\Theta$ in terms of the central density :
\begin{equation}
r_0  = { c_{\rm eff}    \over \sqrt {4 \pi G  \rho_c }  } = { c_{\rm eff}    \over \sqrt {4 \pi G  \chi \rho_{\rm edge} }  }
\end{equation}
\begin{equation}
\xi   = r / r_0,
\end{equation}
and
\begin {equation}
\Theta = \rho / \rho_c,
\end {equation}
where $\rho_{\rm edge}$ is the edge density of the core, and $\chi = \rho_c / \rho_{\rm edge}$ is the density contrast between the center of the core and its edge. Combining the Poisson equation with the momentum equation, assuming the pressure  support is derived from a polytropic equation of state, and demanding both spherical symmetry and force balance in the radial direction, gives us the Lane-Emden equation :
\begin {equation}
{1 \over \xi^2} {d \over d\xi} \left ( \xi^2 {d \Theta \over d \xi} \right) + \Theta^n
= 0
\label {laneemden}
\end {equation}
where $\gamma = 1 + 1/ n$, and $\gamma$ is the ratio of specific heats. We take the boundary conditions at $\xi = 0$ as $\Theta (\xi = 0) = 1$ and $\Theta' (\xi = 0) = 0$, appropriate for centrally condensed initial conditions. While there is no closed-form solution to (\ref {laneemden}) for $\gamma = 1$, we utilize a simple analytic approximation originally suggested by \citet{natarajanlyndenbell97}:
\begin{equation}
\Theta (\xi) =  \left( {A \over C^2 + \xi^2} - {B \over D^2 + \xi^2} \right )
\end{equation}
where the parameters $A, B, C,$ and $D$ are numerical constants parameterizing the solution. \citet{natarajanlyndenbell97} suggest $A = 50$, $B = 48$, $C = \sqrt {10}$, $D = \sqrt {12}$. We assume the density distribution  is critically stable, with a contrast $\chi \simeq 14$ \citep {bonnor57, ebert55}. The resultant approximation accurately describes the critical Bonnor-Ebert sphere to within $1\%$ at all spatial points.

In the absence of turbulence, a critically stable core will have the 
familiar Bonnor-Ebert mass and radius \citep {bonnor57, ebert55} :
\begin {equation}
M_{\rm th} = 1.18 {c_{\rm iso}^3 \over \sqrt {G^3 \rho_{\rm edge} } }
\end {equation}
\label {eqn:thmass}
\begin {equation}
R_{\rm th} = .485 {c_{\rm iso} \over \sqrt {G \rho_{\rm edge} } }
\end {equation}
\label {eqn:thradius}
Where $\rho_{\rm edge}$ is the edge density of the core.

Using the effective soundspeed in place of the isothermal soundspeed, we have expressions for the turbulently-supported Bonnor-Ebert mass and radius: 
\begin {equation}
M_{\rm BE} = 1.18 {c_{\rm eff}^3 \over \sqrt {G^3 \rho_{\rm edge} } }
\end {equation}
\label {eqn:bemass}
\begin {equation}
R_{\rm BE} = .485 {c_{\rm eff} \over \sqrt {G \rho_{\rm edge} } }
\end {equation}
\label {eqn:beradius}

Larson's mean-density law states that the mean density within a given volume scales as the inverse size of that region \citep {larson81}. Hence the column density through a given region should remain roughly independent of the size of the region. Applying this mean-density scaling to our $M_{\rm BE}$ and $R_{\rm BE}$, we find
\begin {equation}
{M_{\rm BE} \over R_{\rm BE}^2} \propto {{ c_{\rm eff} \sqrt {\rho_{\rm edge} \over G}  = {\rm const.}}  }
\end {equation}
Hence, because the effective soundspeed $c_{\rm eff}$ must scale with the turbulent linewidth $\mathcal {M}$, we find that the edge density of the cloud core cannot remain a constant, but must scale as 
\begin {equation}
\rho_{\rm edge} = {\rho_0 \over \sqrt{ 1 + (\mathcal {M} / 3)^2 } }
\end {equation}
where $\rho_0$ is the central density, or equivalently, the edge density as one approaches $\mathcal {M} \rightarrow 0$. In the remainder of the paper, we will set $\rho_0 = 5 \cdot 10^{-20}$ gm cm$^{-3}$, and $c_{\rm iso} = 2 \cdot 10^{4}$ cm s$^{-1}$, which imply $R_{\rm BE} = .07$ pc and $M_{\rm BE} = 2 M_{\odot}$ for a transonic ($\mathcal {M} = 1$) molecular cloud core, which is typical for cores in the Taurus region \citep {jijinamyersadams99}. We note that in a large-scale, isothermal, supersonic molecular cloud, shocks will compress and rarefy the mean flow of the gas, producing a log-normal distribution in the density field, where the width of the distribution depends on the Mach number of the region \citep {padoannordlund02}. In the current work, we treat only the median density, and not the tails of the density distribution. As a result, the minimum mass binary system treated will be proportional to the thermal Jeans mass (see \S \ref {methodology}). This approximation is justified for systems whose masses are each close to the median stellar mass, but above the substellar limit. We will extend the semi-empirical method to systems containing at least one substellar component in a future paper \citep {fisher03}. 

\subsection{Velocity Field Model}
\label{velocityperturbations}

As we noted earlier (\S \ref {velocitystructure}), Larson's linewidth-size relation, as determined by recent authors \citep {jijinamyersadams99} states that the velocity dispersion $\Delta v$ along a  line of sight through a region of size $R$ scales roughly as $R^{1/2}$ on molecular cloud core scales and above.

We assume that each Fourier mode in the velocity decomposition is uncorrelated. By the Central Limit Theorem, uncorrelated Fourier modes are equivalent to demanding that the probability distribution of the amplitude of each mode is drawn from a Gaussian distribution, and the phase of each mode is drawn from a  uniform distribution. Decomposing an individual mode $\delta_{k}$ in Fourier space  in terms of its amplitude and phase as $\delta_{k} = r_{k}  \exp (i \phi_{k}),$ the Gaussian probability distribution $g_{k}$ for the  amplitude $r_{k}$ and phase $\phi_{k}$ can be written as
\begin{equation}
g_{k} (r_{k}, \phi_{k} ) dr_{k} d\phi_{k} =  {2 (r_{k} 
dr_{k} ) \over \sigma_{k} } \left ( {d\phi_{k} \over 2 \pi }\right) 
\exp \left ( {- r_{k}^2 \over \sigma_{k}^2 }\right )
\end {equation}

The standard deviation $\sigma_{k}$ is referred to as the power spectrum of the perturbation spectrum. As is conventional, we write $P (k) = \sigma_k $. By examining the successive moments of a Gaussian distribution, one  can show that only the first two moments -- the mean and standard deviation -- are  needed to entirely specify the distribution. Hence, specifying the mean and  standard deviation  completely determines the Gaussian spectrum. In the following discussion, we assume that the mean of the perturbing field is zero, and that the  perturbing field is homogeneous and isotropic, so that the amplitude of the power spectrum depends solely on the magnitude of the $k-$vector.

By Parseval's Theorem, the total power introduced in $k$-space is identical to the total power in real space. Hence, the 3D rms velocity $\Delta v$ in a 3D spherical volume in real space extending from $L_{min}$ to $L_{max}$ in the radial  direction can be easily evaluated by integrating the power spectrum in k-space from  $k_{min} = 2 \pi / L_{\rm max}$ to $k_{\rm max} = 2 \pi / L_{\rm min}$:
\begin {equation}
(\Delta v)^2 = \int^{4 \pi}_0 d\Omega \int^{k_{\rm max} }_{k_{\rm min} }
 k^2 d {k} P ( k )
\end{equation}
Adopting a power-law power spectrum of the form $P (k) = A k^{-n}$, then  integrating over angles and wavenumbers, we have
\begin{equation}
(\Delta v)^2 =  { 4 \pi A \over (3 - n) \left ( {k_{min}}^{-n +3} - {k_{max}}^{-n + 3}  \right ) }
\end {equation}
Assuming that most of the power is on large scales ($n> 3$) and that $L_{min} \ll L_{max}$, the maximum wavenumber ${k_{max} }$ corresponding to  the minimum scale in the problem can be neglected, so we have
\begin {equation}
\Delta v =  \left (4 \pi A \over n - 3 \right )^{1/2} \left (2 \pi L_{max}^{(-
n +3) / 2} \right )
\end {equation}
When $n = 4$, we can recover Larson's Law ($\Delta v \propto L_{max}^{1 / 2}$). This  $n = 4$ Gaussian velocity perturbation power spectrum provides the basis for our turbulent models. Specifically, we assume all background velocities are  zero, and apply three separate turbulent velocity fields to our model molecular cloud  core via
\begin {equation}
v_x (\vec {r} )  = {\delta_x (\vec {r} )  c_{\rm iso} \mathcal {M} \over \sqrt {3} }
\end{equation}
\begin{equation}
v_y (\vec {r} )  = {\delta_y (\vec {r} )  c_{\rm iso} \mathcal {M} \over \sqrt {3} }
\end{equation}
\begin{equation}
v_z (\vec {r} )  = {\delta_z (\vec {r} )  c_{\rm iso} \mathcal {M} \over \sqrt {3} }
\end{equation}
where $\delta_x (\vec {r} )$,  $\delta_y (\vec {r} )$, and  $\delta_z (\vec {r}
)$ are three realizations of  a $k^{-4}$ Gaussian field, $c_{\rm iso}$ is the isothermal
soundspeed, and $\mathcal {M} = \Delta v / c_{\rm iso}$ is the 3D rms turbulent Mach number. 

While Gaussian perturbation fields and the structures which arise from them have been a topic of active investigation  in the cosmological context for several  decades, our use of them in the context of star formation differs in two critical respects. First, application of a perturbation to the density field requires that the total power must be small : a nonlinear perturbation with sufficient
total power will drive the density negative in some regions. By applying  perturbations to the velocity field directly, the total power can be made  arbitrarily large without encountering unphysical negative densities. Second, whereas cosmological density fluctuations typically introduce most power on small scales (the inflationary Harrison-Zel'dovich spectrum uses $n = 1$), we introduce most power on large scales, in accord with observations of star forming regions.

One should appropriately ask how the ratio of rotational to gravitational binding energy, $\beta$, depends on the 3D turbulent Mach number $\mathcal{M}$, while keeping the realization of the turbulence fixed. Interestingly enough, under the assumption of a critical Bonnor-Ebert density distribution, for transonic and supersonic cores, a fixed realization yields a $\beta$ that is completely independent of the edge-density of the core, and only weakly dependent on the turbulent Mach number. Specifically,
\begin{equation}
\beta \propto \left(J^2 \over I \right) \left (R \over M^2 \right)
\end{equation}
\begin{equation}
\beta \propto \left(\mathcal {M}^2 R \over M \right)
\end{equation}
\begin{equation}
\beta \propto  {\mathcal {M}^2 \over { 1 + \left  ( {\mathcal {M} \over \sqrt {3
} }  \right)^2  } }
\end {equation}
ie, for transonic and supersonic cores with $\mathcal {M} \ge 1$, $\beta$ varies by roughly a factor of 3 from $\mathcal {M} = 1$ to $\mathcal{M} = \infty$. Here $I$ is the moment of inertia of the initial core. This scaling explains why turbulent core models naturally produce the same median value of $\beta$ as is seen in observation without any fine-tuning of parameters: in critical Bonnor-Ebert spheres dominated by turbulent pressure support, Mach scaling applies, and as a result the models are scale-free. Therefore, in the supersonic regime, for a fixed turbulent realization, $\beta$ does not depend on ${\it any}$ model parameters. This scaling holds approximately even down to the transonic regime, so that the value of $\beta$ is primarily determined from the slope of the turbulent spectrum. The robust agreement between the predicted value of $\beta$  and observation gives strong support for our model of the turbulent velocity field.

Similar Gaussian turbulent spectra imposed on the velocity field have been used in a variety of numerical simulations of turbulence in the interstellar medium. \citet {dubinskietal95} were apparently the first to suggest that Larsonian  turbulence in the interstellar medium could be generated using Gaussian random  velocity fields with index $n = 4$. A number of authors, beginning with \citet {gammieostriker96},  studied the evolution of turbulence in molecular clouds using  the same turbulent spectrum -- initially establishing an  Alfv{\' e}nic spectrum of waves by an initial velocity fluctuation on a constant density and magnetic  field background. \citet {burkertbodenheimer00} computed linewidth gradients induced by turbulent eddies on the scale of molecular cloud cores.  \citet{hujeiratetal00} computed the initial decay of turbulence and subsequent collapse (followed to an early evolutionary time) of  molecular cloud cores supported by Alfv{\' e}nic turbulence. Lastly, \citet {bateetal02} computed the evolution of a large, highly supersonic ($\mathcal {M} \sim 7$) turbulent core.

\section {Semi-Empirical Model for Binary Formation via Turbulent Fragmention}
\label {methodology}

We may now predict the distribution of binary periods implied by our isolated turbulent core models. Since our knowledge of the detailed physics of turbulent fragmentation is still rudimentary at best, in order to predict binary properties from our initial molecular cloud core models, we must make several semi-empirical assumptions motivated by observation.  We will assume that the star-formation efficiency $\epsilon_* = (M_1 + M_2) / M_{core}$ 
is constant. Further, in direct analogy to the star-formation efficiency, we define the conversion efficiency of angular momentum of cores to binaries, $\epsilon_J = J / J_{core}$, and assume it is constant as well. In addition, we will assume that the individual stellar masses are uncorrelated, so that both may be randomly drawn from the IMF. Lastly, we will assume that the orbital eccentricities of all binaries are drawn from a distribution $f (e) = 2 e$. 

We recognize that these assumptions may not be rigorously correct; in particular, based on observations of field stars,  there is evidence \citep{dm91} that short ($P < 10^3$ d) period binary systems follow a different eccentricity distribution than long period binaries, and may also have a different mass-ratio distribution \citep {mazehetal92}. However, because such short-period systems will be strongly affected by disk-star interactions, it is likely that their orbital parameters undergo significant evolution. Hence it, in the context of obtaining initial binary properties in the current work, we will explore the consequences of the hypothesis that {\it all} binary systems are formed with uncorrelated masses drawn from the IMF, and obeying a thermal eccentricity distribution.

We proceed as follows.

1) First, we randomly draw two masses $M_1$ and $M_2$ from the IMF. We uniformly draw $\xi$ over the interval from 0 - 1 twice, and assign masses (scaled to solar) and mass ratio $q$ according to
\begin {equation} 
m (\xi) = .08 + {\gamma_1 \xi^{\gamma_2} + \gamma_3 \xi^{\gamma_4} \over (1 - \xi)^{.58} } ,
\end {equation}
where $\gamma_1 = .19$, $\gamma_2 = 1.55$, $\gamma_3 = .050$, $\gamma_4 = .6$ \citep {kgt91}. We note that this IMF introduces a cutoff at $.08 M_{\odot}$, and so does not extend into the substellar range. We define  $M_1$ to be the greater of these two masses; hence $q = M_2 / M_1$.

2) Next, using the star formation efficiency factor, set the mass of the initial protostellar core:
\begin {equation}
M_{\rm core} = {(M_1 + M_2) \over \epsilon_*}.
\end {equation}
Since we now know the parent core mass  as well as the thermal Bonnor-Ebert mass, we may also set the 3D turbulent Mach number, using equation (\ref{eqn:bemass}) :
\begin {equation}
\mathcal {M} = 3 \left[ \left( {M_{\rm core} \over M_{\rm th}} \right)^{4/7} - 1 \right]^{1/2}
\end {equation}  
In cases where $M_{\rm core} < M_{\rm th}$, the computed core mass is less than the thermal Bonnor-Ebert value, implying that the initial core is inconsistent with the selected binary masses and assumed star formation efficiency. In these instances, we simply reject the model binary and draw both stars again.

3) We then stochastically generate three perturbation cubes, and perturb the velocity field of our model core using the known Mach number $\mathcal {M}$, thereby setting the resultant core angular momentum $J_{\rm core}$ as well as the binary system angular momentum $J = \epsilon_J J_{\rm core}$.

4) Next, we draw the eccentricity of the binary orbit from the thermal distribution $f (e) = 2 e$. We do this by drawing a number $\xi$ from a uniform distribution ranging from 0 to 1, and setting $e = \sqrt {\xi}$.  

5) Finally, knowing $J$, $e$, $M$ and $q$, we may compute the period $P$ and semi-major axis $a$ of the binary system, from
\begin {equation}                                                              
P = \left( {2 \pi \over G^2} \right) \left( {J^3 \over M^5} \right) {1 \over (1 - e^2)^{3/2} } { (1 + q)^6 \over q^3}
\label {eqn:binaryperiod1}
\end {equation}   
and
\begin {equation}
a = {1 \over G} \left({J \over M^2} \right) {1 \over (1 - e^2)} {1 \over M} {(1 + q)^4 \over q^2}
\label {eqn:binarya}
\end {equation}
(See Appendix A for derivation.)

These relations are often written down for the special case of an equal-mass, circular binary ($e = 0$, and $q = 1$). Clearly, in that case, the larger the value of  $J / M$, the greater the amount of angular momentum in the system, and hence, the longer the period. In addition, the smaller the
 total mass $M$ of the system, the less the influence of gravity, and hence, the wider the binary.  However, although it is commonly not recognized, for fixed mass and angular momentum,  the eccentricity and the mass ratio can also have a substantial impact on the orbital properties of binaries. For instance,  for a fixed angular momentum, the larger the eccentricity $e$ of a system, the wider the binary. Similarly, for a fixed mass, the smaller the mass ratio $q$, the wider the binary. Although apparently trivial, these scaling relationships will have an important bearing on the statistical properties of binary star systems, as we shall soon see. 

Our models introduce two free parameters: $\epsilon_*$ and $\epsilon_J$, the mass and angular momentum star formation efficiencies, respectively. The first parameter is relatively well-constrained by both theory and observation \citep {matznermckee00}. For a given $\epsilon_*$, the $\epsilon_J$ parameter is obtained by requiring that the resultant model binary period distribution median agree with the observed value. In effect, the imposition of this constraint reduces the parameter space to one free parameter, which we take to be $\epsilon_*$. We note, from equation (\ref {eqn:binaryperiod1}), that with the constraint that the model period median must agree with observation, that $\epsilon_J \propto \epsilon_*^{5/3}$. The nonlinearity of the dependence of $\epsilon_J$ on $\epsilon_*$ has a simple interpretation: for low star formation efficiency, a larger amount of mass must be accreted onto the binary, which implies a greater angular momentum loss efficiency due to the larger turbulent linewidth on larger scales.
 
Note that our model implicitly assumes that the mass of the system $M$, the mass ratio $q$, and the orbital eccentricity $e$ are uncorrelated, whereas the angular momentum $J$ (and hence the period $P$) is explicitly dependent upon $M$ : the higher the mass of the binary system, the greater the mass and Mach number $\mathcal {M}$ of the initial core.  These assumptions differ from those adopted by \citet {kroupa95} in his formulation of the inverse binary population synthesis problem, where he assumed that $P$, $q$, and $e$ were uncorrelated. In contrast, in our formulation, the period $P$ depends explicitly on $q$ and $e$ through angular momentum conservation (equation \ref {eqn:binaryperiod1}), which implies that our model predicts non-trivial correlations between $P$ and $q$, and between $P$ and $e$. We return to these points in later sections (\S \ref {eccvsperiod} and \S \ref {qvsperiod}), where we detail the nature of these correlations.

\section {Results}
\label {results}

\subsection {Binary Period Distribution}

\begin {figure}
  \plottwo{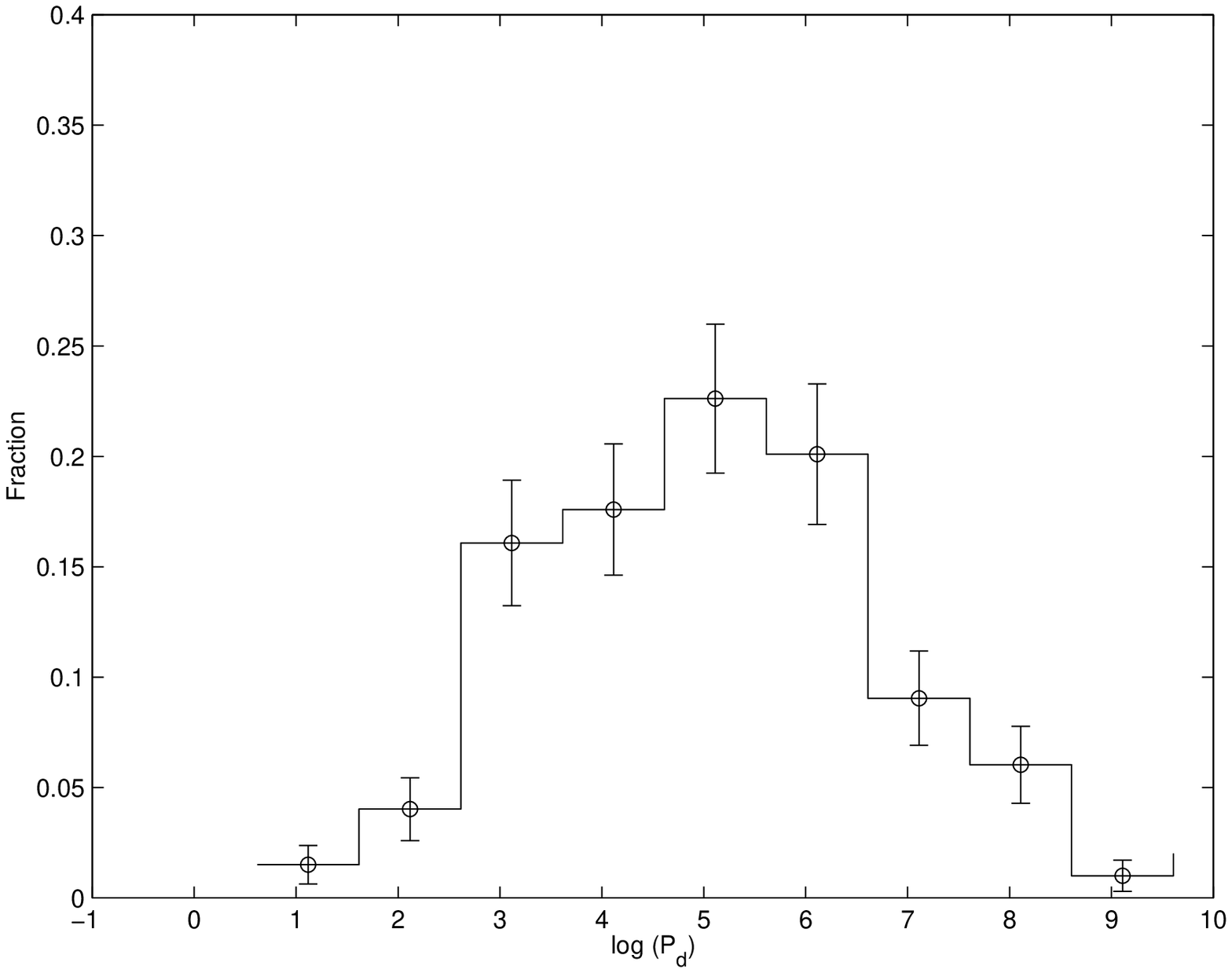}{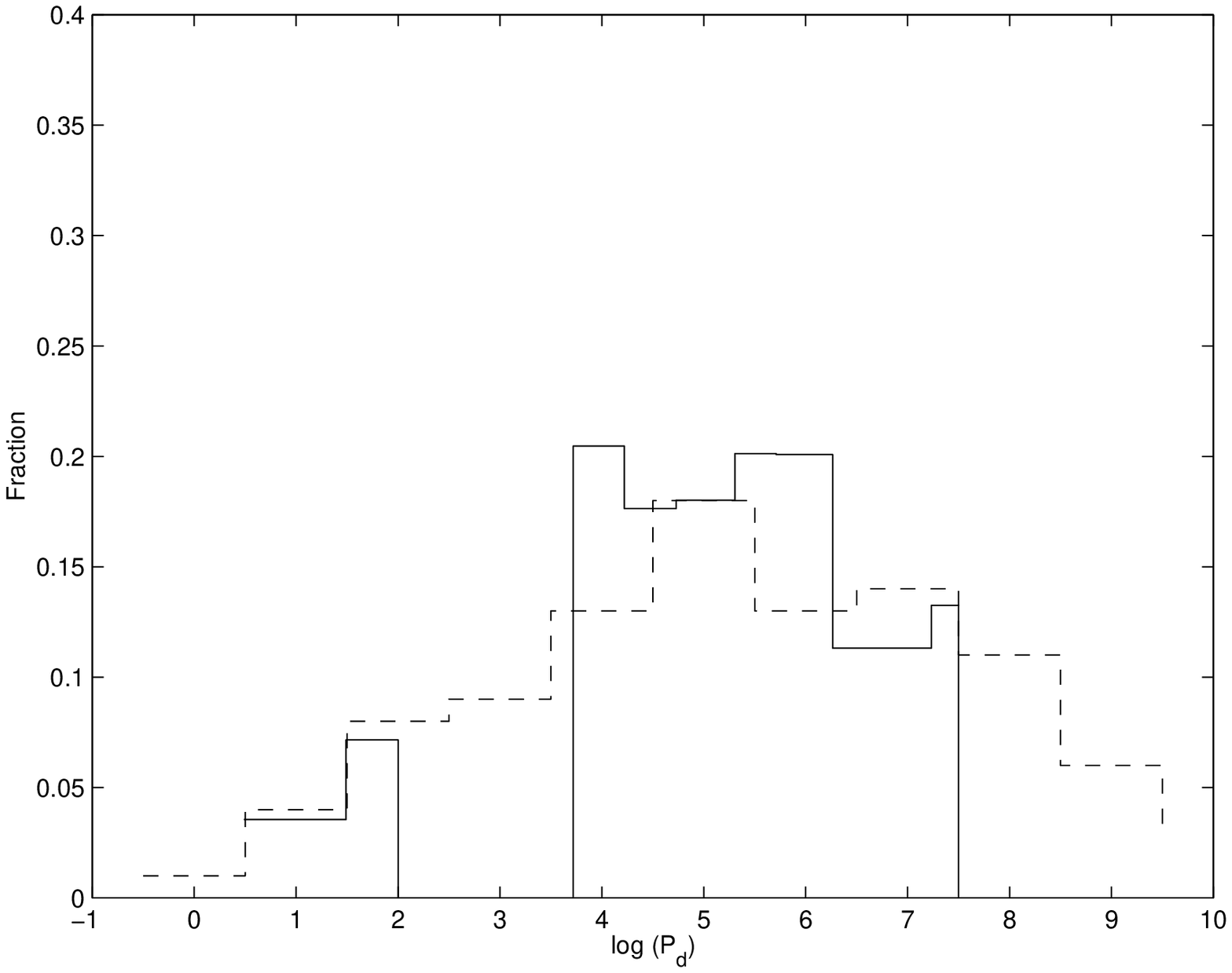}
  \caption{Histogram of $\log P_d$. The left panel shows our numerical results
for 200 model systems for the case of star-formation efficiency $\epsilon_* = 0
.26$.  For comparison, the right-hand panel shows the period distribution inferred
from PMS stars \citep {mathieu94} and from  field stars
 \citep {dm91}, in solid and dashed lines, respectively.}
  \label  {fig:histlogperiod}
\end {figure}

Figure \ref {fig:histlogperiod} shows the binary period distribution obtained for the star formation efficiency $\epsilon_* = 0.26$. The standard deviation of the computed distribution is $\sigma_{P_d} = 1.7$; slightly narrower than that of the field, though consistent with the PMS distribution. The computed distribution is in remarkably good agreement with Mathieu's cataloged  visual binaries with periods $> 10^4$ d. The computed distribution exhibits a somewhat smaller binarity fraction at periods shorter than $10^3$ d than either the field or the PMS distribution, although given the numbers involved, this result is of marginal statistical significance.  

\begin{figure}
  \plotone {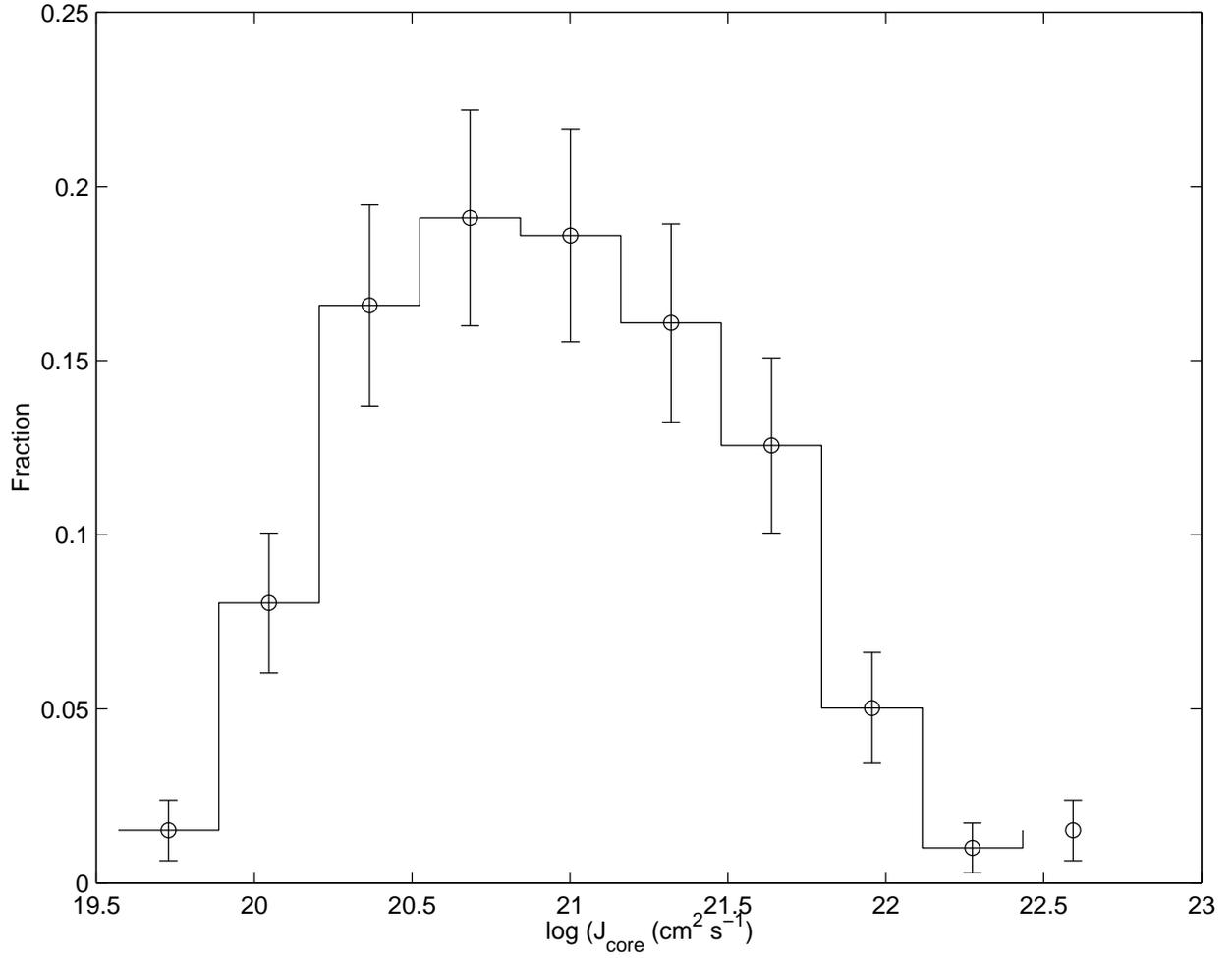}
  \caption {Histogram of the log of the model initial core specific angular mom
entum $j = J / M$, measured in cm$^2$ s$^{-1}$, for 200 binary-producing cores,
 shown for a low star-formation efficiency ($\epsilon_* = 0.26$).}
  \label{fig:histlogspecjcore}
\end{figure}

There is a small, but systematic increase in the width of the distribution with increasing star formation efficiency. We interpret this increasing width as a consequence of the decrease in angular momentum loss efficiency with increasing star formation efficiency. As a result, higher star formation efficiency models have slightly wider distributions of pre-stellar core specific angular momenta. The distribution of initial core specific angular angular momenta for the case of $\epsilon_* = 0.26$ is shown in figure \ref {fig:histlogspecjcore}.

\subsection {Eccentricity Versus Period}
\label {eccvsperiod}

As we noted in \S \ref {methodology}, in our semi-empirical model of turbulent star formation, the angular momentum of the binary system is directly related to the initial angular momentum of the turbulent core. As a result, the period and eccentricity are no longer uncorrelated quantities, and the period is explicitly dependent upon the eccentricity; for a fixed angular momentum and mass, the greater the eccentricity, the longer the period. From observation (\S \ref{eccentricitydistr}), we know that in binaries wider than the tidal circularization limit ($P \sim 10$ d), the period tends to be correlated with the eccentricity, which is in fact a robust feature of our model (see equation \ref {eqn:binaryperiod1} ). 

We plot eccentricity versus the log (base 10) of the period for in figure \ref {fig:evslogp}, for the case of intermediate star-formation efficiency ($\epsilon_* = 0.5$).  For comparison, binaries from the field \citep {dm91}, and from PMS
 regions \citep {mathieu94}, are shown in star and plus symbols, respectively,
in the plot on the right. The vertical dashed line indicates the tidal 
circularization limit found by \citet {dm91} at about $P_d \sim 11$ d. The key result here is that the model systems exhibit a positive correlation between eccentricity and period, which is qualitatively similar to that observed. However, it is certainly true that our models predict an overabundance of highly eccentric binaries at shorter periods. It is quite likely that additional angular momentum dissipation mechanisms, including coupling to circumstellar disks, may help circularize such systems. 

\subsection {Mass Ratio Versus Period}
\label {qvsperiod}

The semi-empirical model also predicts a correlation between mass ratio and period. The reason for the correlation is, once again, simply related to conservation of angular momentum. A system with two unequal masses, and a small mass ratio ($q \ll 1$) will have a longer period than a similar system with the same total angular momentum and mass, but with equal masses. The results of 200 model systems for the case of intermediate star-formation efficiency ($\epsilon_* = 0.5$) are shown in figure \ref {fig:qvslogp}. 

We emphasize that in order to explain the correlation between mass ratio and period, the semi-empirical model framework {\it does not require that the masses themselves be correlated}. Instead, the model suggests that it is possible to form two stars independently within a single turbulent core, with the resultant period anti-correlated with the mass ratio simply through angular momentum conservation. 

\subsection {Binary System Mass Versus Semimajor Axis}
\label {binarymassvsa}

As we noted previously, observations of low-mass systems have revealed that such systems do not follow the same statistical trends observed in solar-mass and higher binaries. Perhaps the most noticeable difference is the trend for low-mass systems to occur in tighter binaries. 
In the semi-empirical model, two factors contribute to the explanation of this effect. First, lower-mass binary systems originate from parent cores of lower mass, and hence, lower specific angular momentum. Second, drawing stars independently from the IMF, a low-mass system will naturally tend to have a more equal-mass ratio.  Taken together, the model predicts that low-mass systems should naturally tend to occur in shorter-period systems, which is in fact what is observed \citep {martinbasri01}. 

In figures \ref {fig:logavslogmtot_lowsf} -  \ref {fig:logavslogmtot},
we display the results from two model systems of variable star formation efficiency. The minimum system mass represented for a given star-formation efficiency $\epsilon_*$ is indicated with a dashed horizontal line; no model systems form below the line. For comparison, we also plot the largest observable binary separations $a_{\rm max}$, as determined by \citet {closeetal03} in their figure 15. They found that the two straight lines of constant orbital velocity $a_{\rm max} = 23.2 (M_{\rm tot} / .185)$ AU and $a_{\rm max} = 1000 (M_{\rm tot} / .185)$ AU (where $M_{\rm tot}$ is in solar masses) describe the maximal separation upper envelopes relatively well. 

We note that, above the horizontal minimum mass line, our model systems fit the Close et al. upper envelopes remarkably well; very few systems lie to the right of the envelopes, quite independent of the star-formation efficiency. In the context of the current paper, our assumption of a fixed Jeans mass requires that low mass systems must form in regions of low star-formation efficiency. Hence, it appears that a combination of a variety of star-formation efficiencies can explain the results of Close et al. More general models including constant star-formation efficiency and supersonic turbulence on large scales may also be able to explain the observations; a topic we are investigating further \citep {fisher03}.  

\section {Discussion}
\label {discussion}

One limitation of our current approach is the choice of a fixed central density $\rho_0$. 
By specifying $\rho_0$, we have implicitly restricted our attention to systems above a mass $\epsilon_* M_{th}$. As we pointed out previously (\S \ref {backgrounddensity}), shock compressions in a turbulent molecular cloud will produce a log-normal distribution of $\rho_0$. We expect that a set of binary systems formed from cores with varying $M_{th}$ resulting from to shock compression will in fact have a slightly broader period distribution than that computed here. Similarly, the IMF adopted introduces a cutoff at $.08 M_{\odot}$, even though substellar objects are also formed in star-forming regions. These effects may partially account for the slightly smaller period distribution standard deviations computed here ($\sigma_{P_d} \simeq 1.6 - 2.1$), in comparison to that found in the field ($\sigma_{P_d} \simeq 2.3$) \citep {dm91}.  

  Detailed theoretical work treating wind-driven outflows \citep {matznermckee00} predicts that the star-formation efficiency on the molecular cloud scale should lie in the range 25 - 70\%. For star-formation efficiencies in this range, our models predict core properties consistent with observation. In particular, with our fiducial scaling, the median Mach number $\bar \mathcal {M}$ lies in the range of 1.2 and 0.9 over the range $\epsilon_* = 0.1
 - 0.9$  -- quite close to the actual median value of .9 observed in Taurus \citep {jijinamyersadams99}. We note that since both our pre-stellar core and final binary properties are relatively insensitive to the star-formation efficiency (the median prestellar core Mach number  $\bar \mathcal {M}$ varies only by a factor of 2 for star-formation efficiencies over the range $\epsilon_* = 0.1 - 0.9$) our models do not provide further constraints on the star-formation efficiency itself. 

A key consequence of the semi-empirical model for star formation described here stems from the fact that many observational properties can be quantitatively described with a {\it constant} angular momentum loss factor. In contrast, several authors have advocated a ``chaotic" description star formation \citep {larson01, bateetal02}, in which a variety of cluster interactions play a key role in setting the initial properties of binaries. In this paper, we have explicitly demonstrated that chaotic interactions are not required to explain many properties of binaries wider than $P > 10^3$ d, provided that turbulent fragmentation can in fact directly produce the mass and eccentricity distributions which we have assumed. Note that we distinguish the fact that our initial conditions are turbulent, and therefore {\it inherently} stochastic (though deterministic) from the sensitivity upon initial conditions implied by a chaotic model. We would, however, agree that dynamical friction is likely to play an important role in the formation of tight binaries, and that subsequent tidal interactions may also act to broaden the initial period distribution found here.   

Another key consequence of our model is the importance which angular momentum loss from initial molecular cloud cores plays in determining the properties of binaries. Although this problem has been identified by a number of authors \citep {simon92, bodenheimer95, larson02}, virtually all simulations of multiple star formation on the molecular cloud core scale done to date remain purely hydrodynamic, and have neglected the influence of the magnetic field entirely.\footnote {To the best of our knowledge, the only exceptions are the work of \citet {hujeiratetal00} and \citet {boss00}, although Boss did not include a fully self-consistent treatment of the MHD equations.} As a result, the only means of angular momentum transport treated in most calculations is gravitational torquing. Given the importance of angular momentum loss in determining binary properties, and the relative effectiveness of magnetic braking in axisymmetric simulations of single-star formation \citep {basumouschovias94}, caution must be applied in comparing the results of purely hydrodynamic simulations directly against observation; in all likelihood, the magnetic field plays a critical role in determining the binary angular momentum. 

\acknowledgements
 
RTF would like to thank Richard Klein and Christopher McKee for their prescient suggestion to investigate the physical role of turbulence in multiple star formation as a thesis topic. Thanks also go to Marc Davis and Matt Craig for the use of their Gaussian perturbation generation program. This research has made use of NASA's Astrophysics Data System Bibliographic Service. 

\appendix

\section{Characteristic Scales}
\label{characteristicscales}

In this appendix, we derive the characteristic scales of a binary system consisting of two masses $M_1$ and $M_2$, in terms of its angular momentum $J$, its mass ratio $q = M_2 / M_1$, and its orbital eccentricity $e$.

From standard two-body classical mechanics in an inverse square law potential \citep{goldstein80}, we know the angular momentum of a binary system is completely specified by the semi-major and semi-minor axes of the bound orbit, as well as the reduced mass of the two-body system $\mu = M_1 M_2 / (M_1 + M_2)$, and the inverse square law constant of proportionality $k$ :
\begin{equation}
J = {b \over a^{1/2}}  \sqrt {\mu k} ,
\end{equation}
where $\mu = M_1 M_2 / (M_1 + M_2)$ is the reduced mass of the two body system, and $k = G M_1 M_2$ for the gravitational force. Hence, since $b = a \sqrt {1 - e^2}$, we can determine the final orbital separation as a function of the eccentricity $e$, total mass $M = M_1 + M_2$, and specific angular momentum $J / M$ :
\begin{equation}
J / M = a^{1/2} \sqrt {1 - e^2} \sqrt { {G (M_1 M_2)^2 \over (M_1 + M_2)^3} } 
\end{equation}
Defining the ratio of secondary to primary mass $q = M_2 / M_1$, for $e \leq 1$, the semi-major axis of a bound orbit is 
\begin{equation}
 a = {1 \over G} \left({J \over M} \right)^2 {1 \over {1 -e^2} } {1 \over M}  {(1 + q)^4 \over q^2} 
\end{equation}
For a typical molecular cloud core specific angular momentum value of $J / M = 3 \cdot 10^{20}~ {\rm cm}^2/{\rm s}$,
\begin{equation}
a = 50 {\rm AU}  \left ( {{J / M} \over  3 \cdot 10^{20}~ {\rm cm}^2/{\rm s} } \right)^2  {1 \over {1 -e^2} } \left ({M_{\odot} \over M}\right)  {(1 + q)^4 \over q^2} 
\label{binarysemimajor}
\end{equation}
Then, from Kepler's third law, we can determine the period of the final binary system :
\begin{equation}
P = {2 \pi \over G^2}{1 \over (1 - e^2)^{3/2}} \left ({J \over M} \right)^3 { (1 + q)^6 \over q^3} {1 \over M^2} 
\end{equation}
\begin{equation}
P = 300~ {\rm years} {1 \over (1 - e^2)^{3/2}} \left ( {{J / M} \over  3 \cdot 10^{20}~ {\rm cm}^2/{\rm s} } \right)^3  { (1 + q)^6 \over q^3} \left ({M_\odot \over M} \right)^2
\label {binaryperiod}
\end{equation}
We anticipate that explicit dependence of $P$ on $e$ in equation (\ref {binaryperiod}) may cause some readers, who are accustomed to the fact that the Keplerian period depends only on $a$, and not on $e$,  some undue consternation. These concerns will be rapidly alleviated when it is recognized that, for fixed masses and angular momentum, a more eccentric binary system must have a wider separation (eqn. \ref {binarysemimajor}). Our equation (\ref {binaryperiod}) expresses the period in terms of the angular momentum $J$ in favor of the semimajor axis $a$, and hence the origin of the explicit dependence of $P$ on $e$. 

Lastly, using the classical two-body inverse square law result for the total energy $E$ of the system
\begin{equation}
E = -  {k \over 2 a},
\end{equation}
one can also determine the final energy $E$ of the  binary system :
\begin{equation}
E = {1 \over 2} G^2 M^3 (e^2 - 1) {q^3 \over (1 + q)^6 } \left ({M \over J} \right)^2 
\label{binaryenergy}
\end{equation}




\appendix




\clearpage


\begin{table}
\begin{tabular} {|l	|l	|l	|l	|l	|l	|l	|l	|} \hline
$\epsilon_*$ &	 $\epsilon_J$ &  $\log {\bar {P}_d}$ & $\log {\sigma_{P_d}}$ &
	$\epsilon_* M_{th} / M_{\odot}$ & $\bar {a} /$AU & $\bar {j}$ (cm$^2$ s$^{-1}$) & 
	$\bar {\mathcal {M}}$ \\ \hline
0.1  & .002& 4.9  & 1.6  & .14 & 30 & $1.3 \cdot 10^{21}$ & 1.7  \\ \hline
0.3  & .01 & 4.9 &  1.7  & .31 & 34 & $7.7 \cdot 10^{20}$ & 1.2  \\ \hline  
0.5  & .04 & 4.8  & 1.8  & .51 & 30 & $4.5 \cdot 10^{20}$ & 1.0  \\ \hline
0.7  & .06 & 4.9  & 2.0  & .72 & 35 & $3.5 \cdot 10^{20}$ & .89  \\ \hline
0.9  & .10 & 5.0  & 2.1  & .92 & 45 & $3.4 \cdot 10^{20}$ & .87  \\ \hline  
 \label{resultstable}
\end {tabular}
   \caption{A table of model parameters, for a variety of star formation efficiencies
($\epsilon_* = 0.1 - 0.9$) fitted such that the median model period agrees
with that of the field. Shown are the mass and angular momentum star-formation 
efficiencies $\epsilon_*$ and $\epsilon_J$, the median of the log of the period
distribution (in days) $\log {\bar {P}_d}$, the standard deviation of the log of the period distribution
(in days) $\log {\sigma_{P_d}}$, the minimum mass binary system (in solar masses) $\epsilon_* M_{th} / M_{\odot}$, the median of the semi-major
axis (in AU) $\bar {a}$, the median prestellar core angular momentum $\bar {j}$ (in cm$^2$ s$^{-1}$), and the median
prestellar core Mach number $\bar {\mathcal {M}}$.}
\end {table}

\begin{figure}
  \plotone{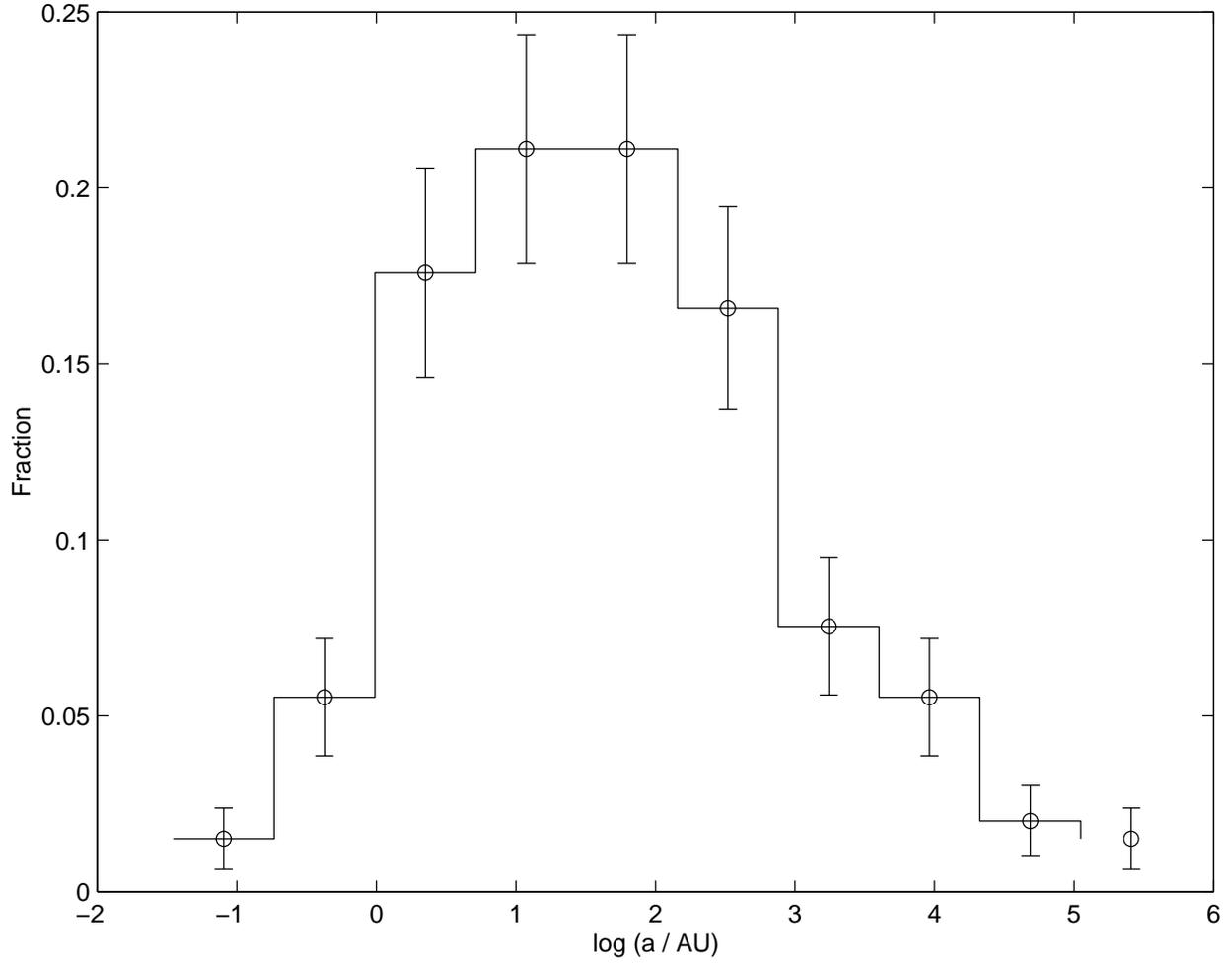}                         
  \caption{Histogram of $\log a$(AU). The numerical results
 for 200 model systems for the cases of low star formation efficiency ($\epsilon_* = 0.26$), are shown, with Poisson error bars drawn.}
  \label  {fig:histloga_lowsf}                                               
\end {figure}

\begin{figure}
  \plotone {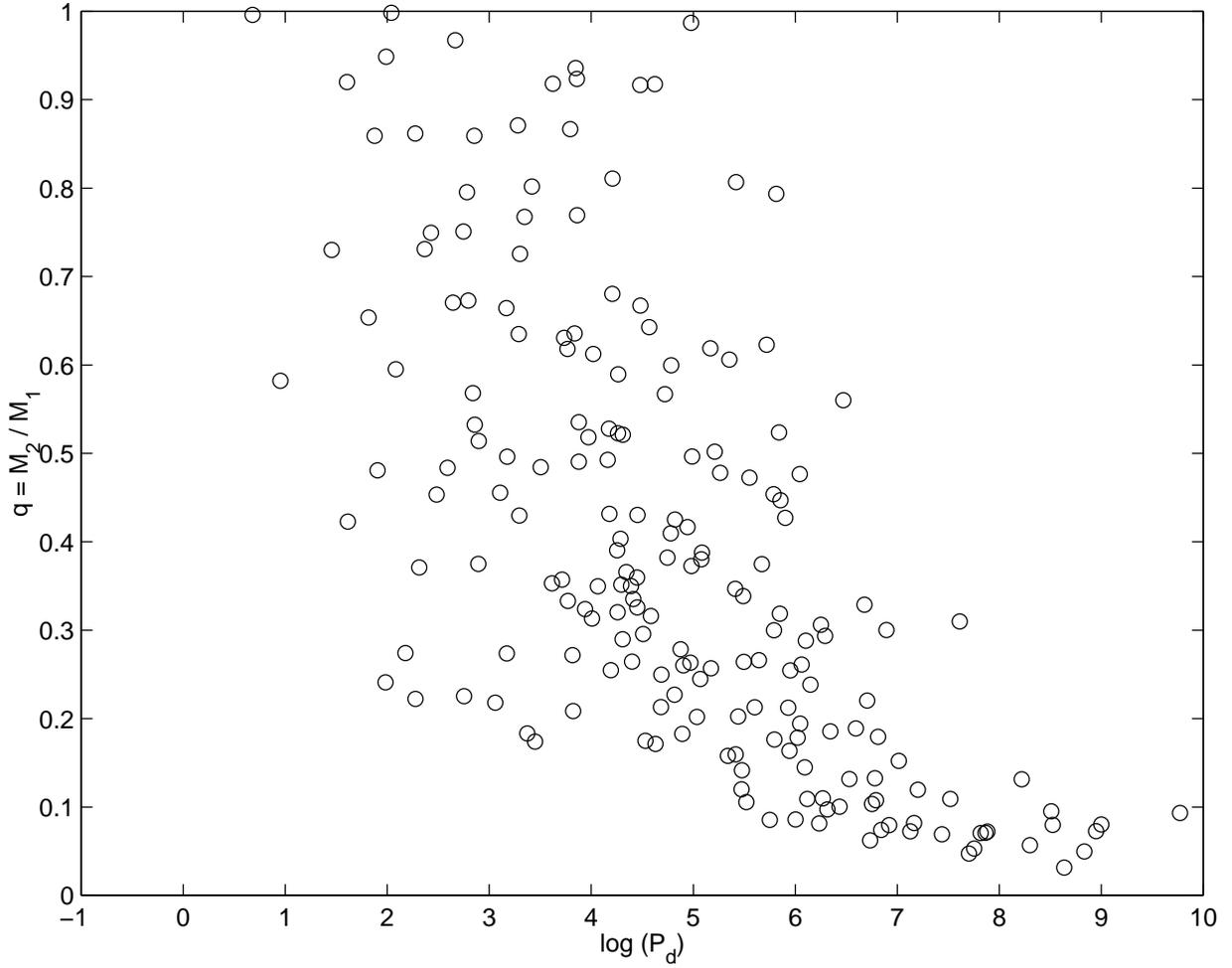}                                 
  \caption{Binary mass ratio versus log $P_d$. The circles show the results of
 200 model systems for the cases of intermediate star-formation efficiency ($\epsilon_* = 0.5$).}    
  \label  {fig:qvslogp}                                                
\end {figure}

\begin{figure}
  \plottwo{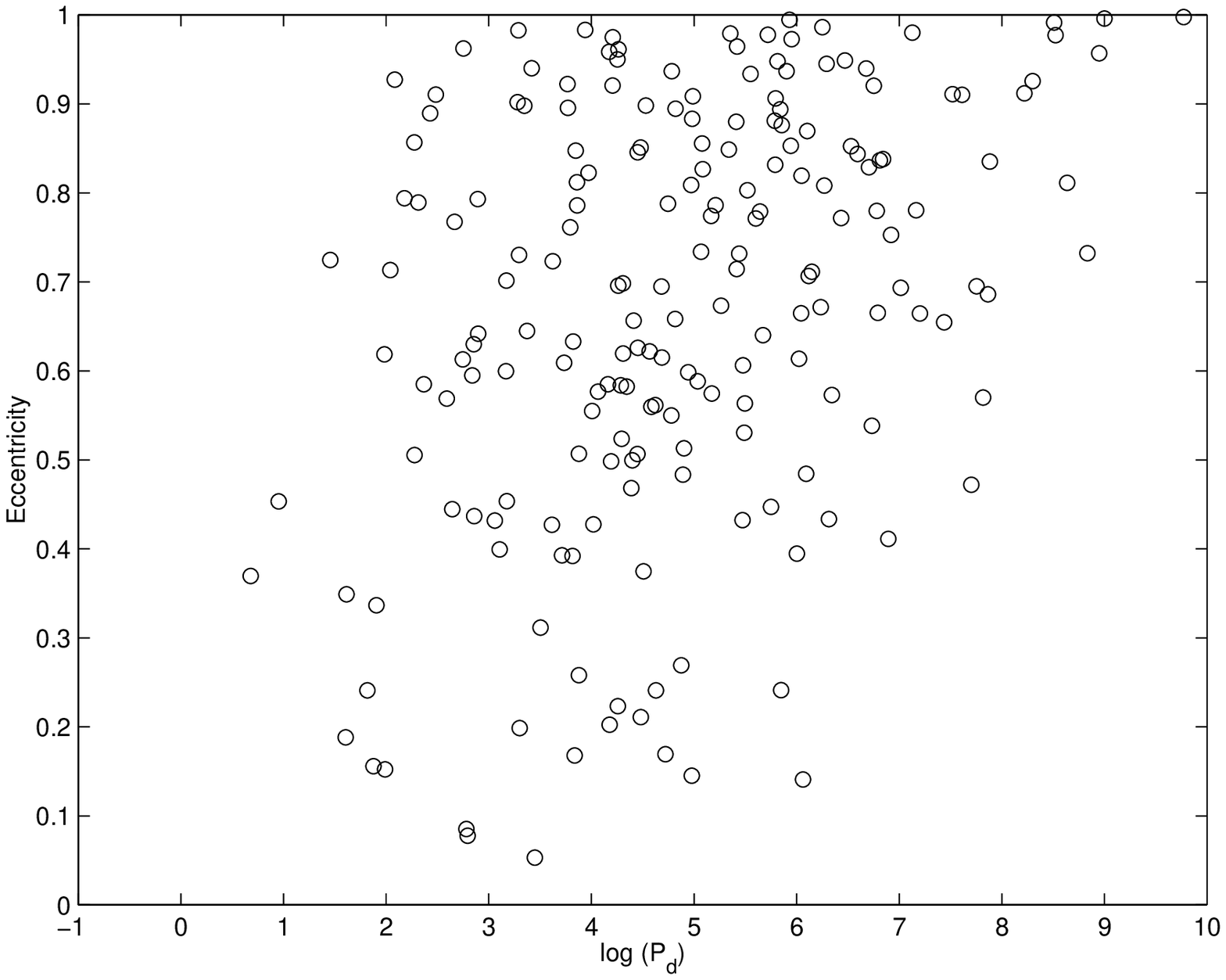} {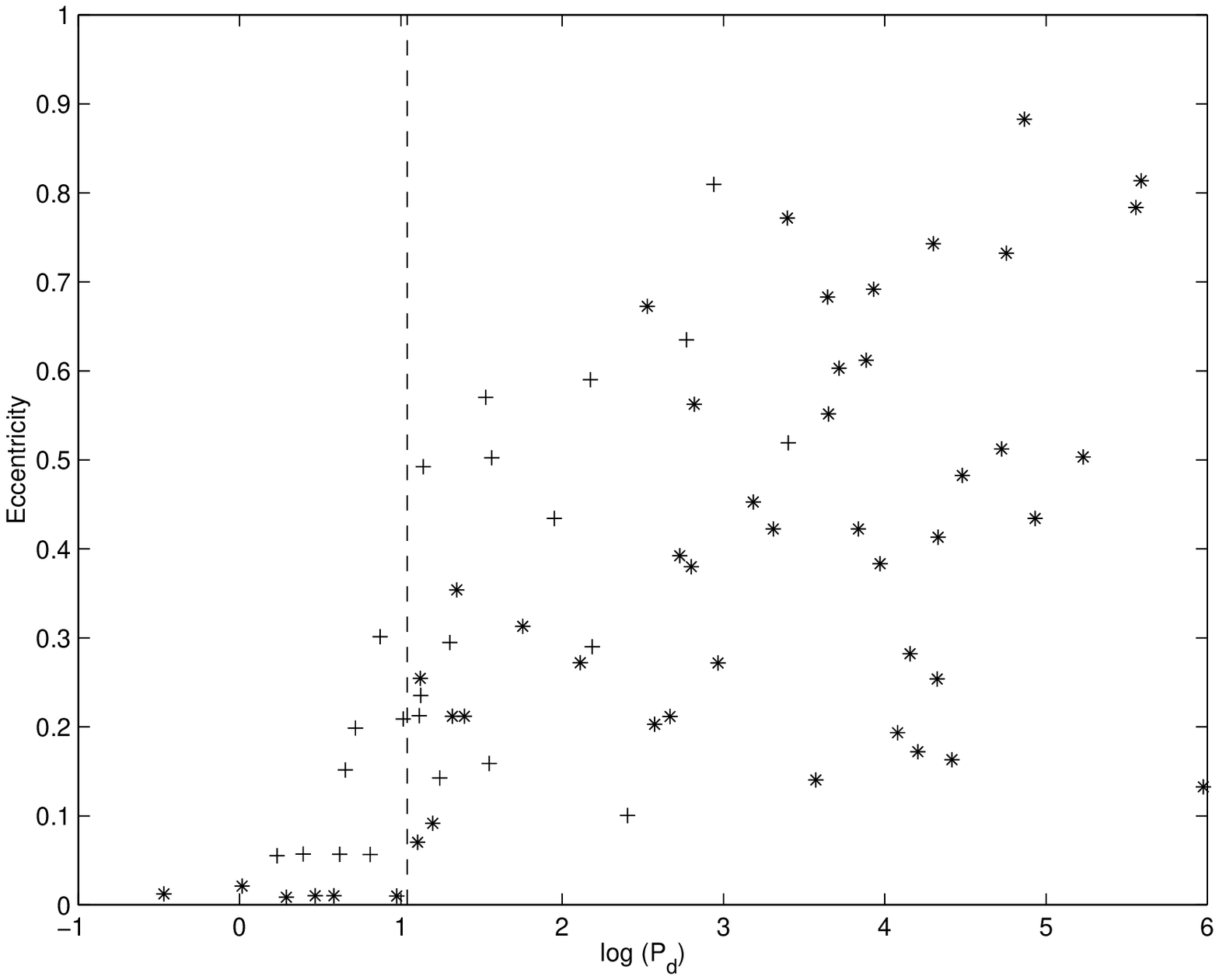}
  \caption {Eccentricity versus log $P_d$. The circles show the results of 200 model systems for the case of intermediate star-formation efficiency ($\epsilon_* = 0.5$). For comparison, binaries from the field \citep {dm91}, and from PMS regions \citep {mathieu94}, are shown in star and plus symbols, respectively, in the plot on the right. The vertical dashed line indicates the tidal circularization limit found by \citet {dm91} at about $P_d \sim 11$ d.}
  \label {fig:evslogp}
\end {figure}

\begin{figure}
  \plotone{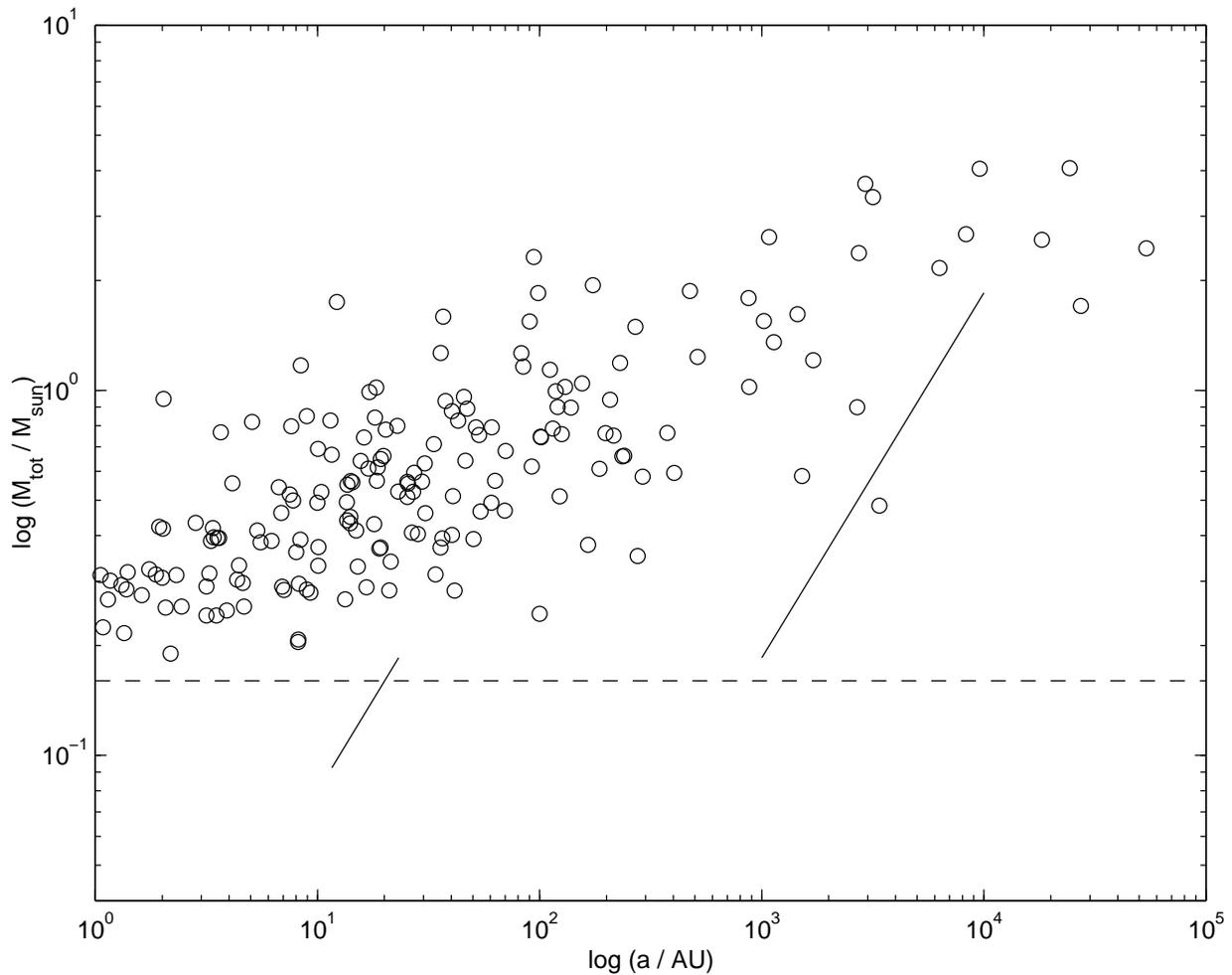}                                                 
  \caption {Log semimajor axis versus log total mass. The circles show the 
results of model systems for the very low star-formation effficiency case $ \epsilon_*
 = 0.1$. The two solid slanted lines indicate the upper envlopes indicated
by \citet {closeetal03} (see text). The horizontal dashed line is drawn to 
delineate the minimum mass binary system for this star-formation efficiency.}
  \label {fig:logavslogmtot_lowsf}
\end {figure}

\begin{figure}
  \plotone{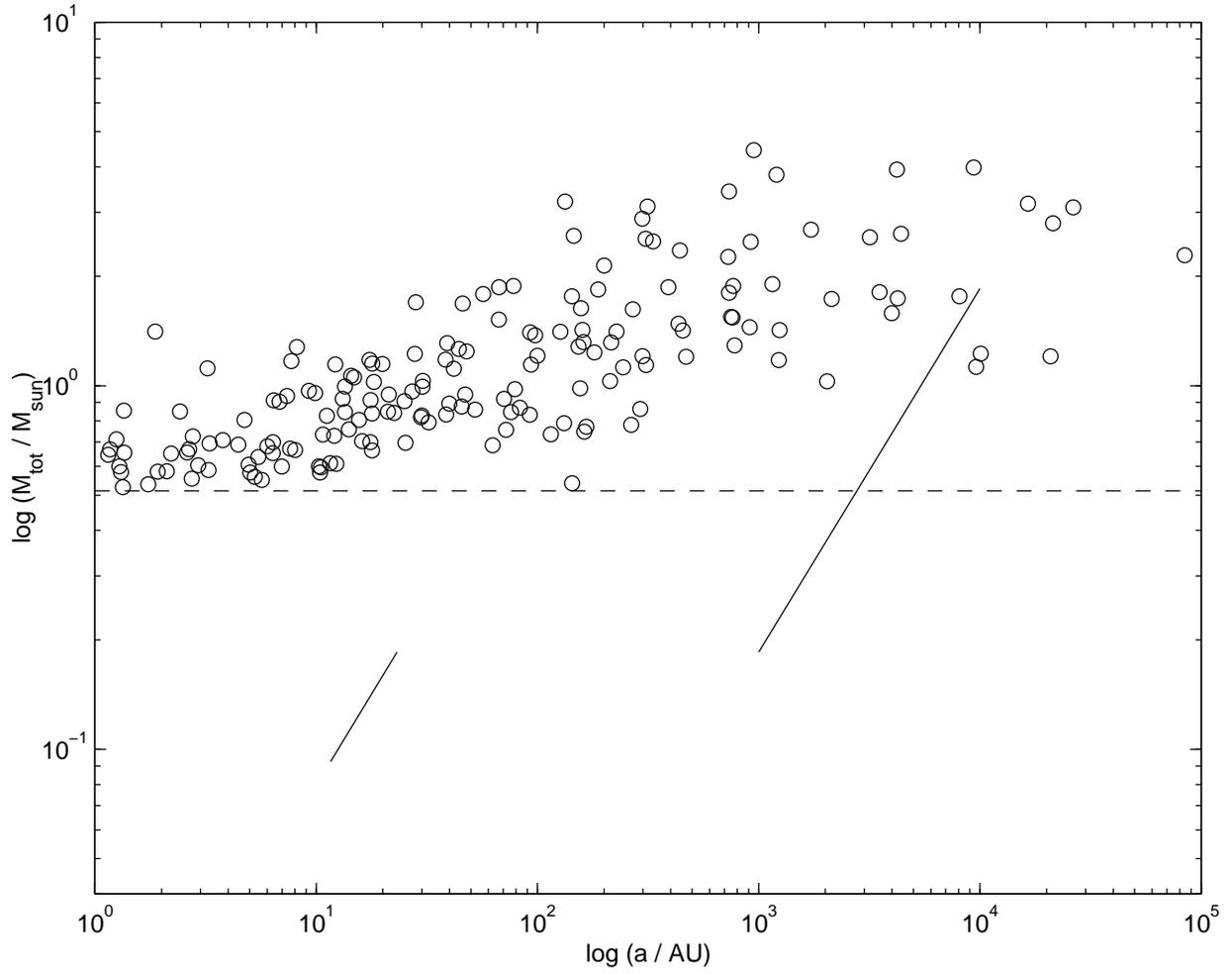}
  \caption {Log semimajor axis versus log total mass, as in fig. \ref {fig:logavslogmtot_lowsf}. The circles show the results of model systems for the intermediate star formation effficiency case $\epsilon_* = 0.5$. }
  \label {fig:logavslogmtot}
\end {figure}

\end{document}